# Ultrashort PW laser pulse interaction with target and ion acceleration

S. Ter-Avetisyan[1], P. K. Singh[2], K. F. Kakolee[2], H. Ahmed[3], T.W. Jeong[2,4], C. Scullion[3], P. Hadjisolomou[3], M. Borghesi[3], and V. Yu. Bychenkov[5,6]

[1]*ELI-ALPS, Szeged 6728, Hungary*
[2]*Center for Relativistic Laser Science, Institute of Basic Science (IBS), Gwangju 61005, South Korea*
[3]*School of Mathematics and Physics, The Queen's University of Belfast, Belfast, BT7 1NN, UK*
[4]*Department of Physics and Photon Science, Gwangju Institute for Science and Technology, Gwangju 61005, South Korea*
[5]*Center for Fundamental & Applied Research, VNIIA, ROSATOM, Moscow 127055, Russia*
[6]*P. N. Lebedev Physics Institute, Russian Academy of Sciences, Moscow 119991, Russia*

**Abstract**

We present the experimental results on ion acceleration by petawatt femtosecond laser solid interaction and explore strategies to enhance ion energy. The irradiation of micrometer thick (0.2 - 6.0 µm) *Al* foils with a virtually unexplored intensity regime ($8\times10^{19}$ W/cm$^2$ - $1\times10^{21}$ W/cm$^2$) resulting in ion acceleration along the rear and the front surface target normal direction is investigated. The maximum energy of protons and carbon ions, obtained at optimised laser intensity condition (by varying laser energy or focal spot size), exhibit a rapid intensity scaling as $I^{0.8}$ along the rear surface target normal direction and $I^{0.6}$ along the front surface target normal direction. It was found that proton energy scales much faster with laser energy rather than the laser focal spot size. Additionally, the ratio of maximum ion energy along the both directions is found to be constant for the broad range of target thickness and laser intensities.

A proton flux is strongly dominated in the forward direction at relatively low laser intensities. Increasing the laser intensity results in the gradual increase in the backward proton flux and leads to almost equalisation of ion flux in both directions in the entire energy range. These experimental findings may open new perspectives for applications.

## 1. Introduction

Laser-driven ion acceleration has shown remarkable advancement in generating the multi-MeV ion bunches and have made realistic progresses in developing laser-driven ion sources as a reliable, generic technology for applications [1]. In recent years, tremendous developments in laser technology has resulted in an access of unprecedented intensities above $10^{21}$ W/cm$^2$ with remarkably improved pulse characteristics, e.g., temporal contrast could be significantly increased by employing several techniques such as cross-polarized wave [2], plasma mirrors [3], saturable absorbers [4]. In this new laser intensity regime, given the novelty of the system and the regime of investigation it is fully appropriate to carry out basic study to obtain / confirm the scaling laws and to test different acceleration mechanisms. This work could be a key for the future development and prospective of ion acceleration with ultrashort drivers.

The acceleration of ions driven by 1.5 petawatt femtosecond laser has been investigated. Irradiation of micrometer thick foils lead to the acceleration of protons and carbon ions along the rear surface target normal (forward) and front surface target normal (backward) direction. The maximum of proton and carbon ion energies showing fast intensity scaling as $I^{0.8}$ along the forward direction and in a backward direction as $I^{0.6}$. It was shown that fast scaling from the target rear $\sim I^{0.8}$ can be attributed to the enhancement of laser energy absorption, as already observed at relatively low intensities. The backwards acceleration of the front side protons with intensity scaling as $\sim I^{0.6}$ can be attributed to the to the formation of a positively charged cavity at the target front via ponderomotive displacement of the target electrons at the interaction of relativistic intense laser pulses with solid target [5]. Furthermore, it was found that the intensity scaling of maximum proton energy is much faster with variation of laser energy for a given laser spot size ($I^{0.8}$), as compared to changing the laser focal spot size for a fixed laser energy ($I^{0.25}$).

The ratio of maximum proton energy along the forward and backward direction is found to remain almost constant for a wide range of target thicknesses (0.2 - 6.0 µm) and laser intensities ($8\times10^{19}$ W/cm$^2$ - $1\times10^{21}$ W/cm$^2$). A symmetric acceleration of protons on both side of the target has been observed in [6] by irradiating the foils with high contrast ($10^{-10}$) laser pulses. However, in this study the intensity was limited to $5\times10^{18}$ W/cm$^2$, where ponderomotive forces does not play a dominant role. In recent experiments [7], no significant acceleration in the backward direction was observed when intensity regime was extended to $2\times10^{20}$ W/cm$^2$.

Additionally, the proton flux along the target normal forward and backward direction is examined by increasing laser intensity. It was found that at low laser intensity proton flux is strongly dominated in the forward direction. An increase in laser intensity results an increase in the ion flux along target normal backward direction. This gradually equalizes with the flux from target normal forward direction in the entire energy range.

## 2. Experimental Set up

The experiment was performed with petawatt, Ti:Sapphire femtosecond laser, installed at the Center for Relativistic Laser Science (CoReLS), Gwangju [8]. A detailed schematic of the experimental design is shown in Fig. 1. A 30 fs, *p*-polarized laser pulse was focused using an f/3 gold-coated off-axis parabolic mirror on (0.2 – 6.0) µm *Al* foils at an angle of 30º. The focal spot, measured with attenuated laser energy, had nearly 30 % of energy confined in the 4 µm FWHM, resulting in maximum laser intensity of $1.2\times10^{21}$ W/cm$^2$. A "saturable absorber" enables a laser pulse temporal contrast, measured by a scanning third order cross correlator, ~$10^{-10}$ in a few picosecond before the main pulse. The nanosecond pedestal was well below $10^{-12}$.

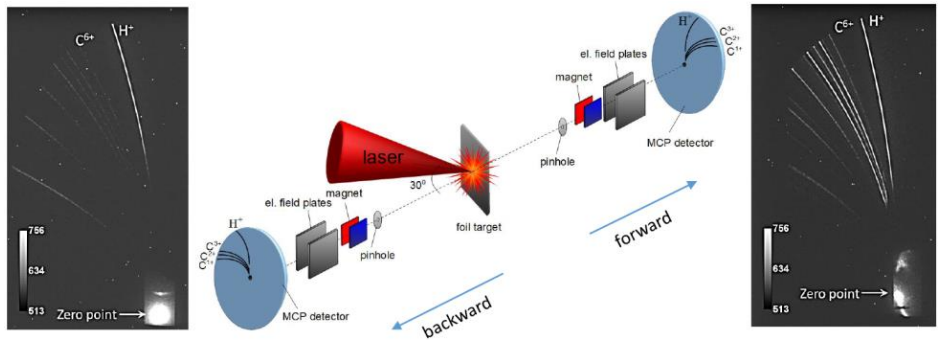

Fig. 1. The laser accelerated ions along the rear surface target normal (forward ions) and front surface target normal (backward ions) direction captured by the Thomson spectrometers. Raw parabolic traces of ions, obtained by irradiating 6 µm thick Al foil at laser intensity of $8\times10^{20}$ W/cm$^2$ are shown as an example.

A laser focal spot and target position monitoring diagnostic system, with an accuracy of few micrometers, was used for precise alignment of the targets at the laser focal plane [9]. Two Thomson spectrometers, with an absolute calibrated microchannel plate (MCP) detector [10], were used to record the energy spectrum of accelerated ions along the rear surface target normal (forward) and front surface target normal (backward) direction. The ion collection solid angle in both spectrometers was $4.67\times10^{-9}$ sr, resulting in clear separation of proton and carbon ion traces on the MCP detector plane. The MCP phosphor screen was imaged using a 16 bit CCD (PIXIS 1024) camera.

Figure 1 shows typical parabolic traces of ions accelerated along the forward (right figure) and backward (left figure) direction, obtained by irradiating 6 µm *Al* foil at laser intensity of $8\times10^{20}$ W/cm$^2$, are shown. No ion emission was observed along the laser propagation axis (30º from target normal), indicating that the particles accelerated along the target normal direction are much collimated. This is also evident from the proton beam profile measured with radio chromic films. The proton divergence estimations put maximum angle of ~ 25º for low energy protons.

## 3. Forward - backward correlation in ion acceleration

Figure 2 a) and b) shows the accelerated protons and $C^{6+}$ ions maximum energies accelerated from 0.2 µm thick *Al* target, measured in arbitrary consecutive laser shots along the forward and backward directions. The ratio of forward and backward cut-off energies (blue diamonds) stays virtually constant despite the dramatic variation of proton and $C^{6+}$ ion energies due to changes in the laser intensity. The ratio for protons stays nearly constant with an average value of 1.5 ± 0.2, and for $C^{6+}$ with an average value of 2.2 ± 0.3. The flat response of the ratio indicates a strong correlation between acceleration of ions on both sides of the targets: for protons and for the much heavier carbon ions.

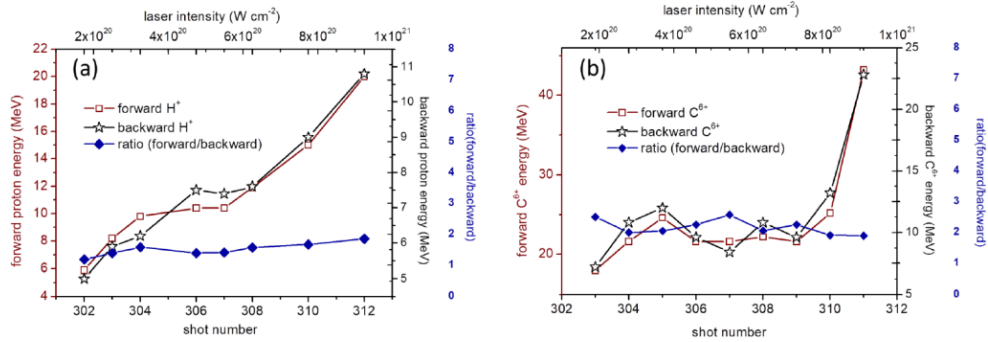

Fig. 2. Forward backward correlation for proton and Carbon ions. Maximum energy of accelerated ions from 0.2 µm thick *Al* target measured in arbitrary consecutive laser shots along the forward and backward direction for a) proton and b) $C^{6+}$ ions where laser intensity is ramped up. The ratio of forward and backward maximum energy is shown by blue diamond.

The proton flux has also been examined for different laser intensities with Fig. 3 a) and b) shows energy spectrum of forward and backward accelerated protons from 0.2 µm *Al* foil irradiated to an intensity of a) $1.5\times10^{20}$ W/cm$^2$, and b) $7.2\times10^{20}$ W/cm$^2$. At low laser intensity (Fig. 3a)), proton flux is strongly dominated in the forward direction in the entire energy range. An increase of laser intensity results in a gradually increase in the backward proton flux which leads to an almost equalisation of ion flux in both direction.

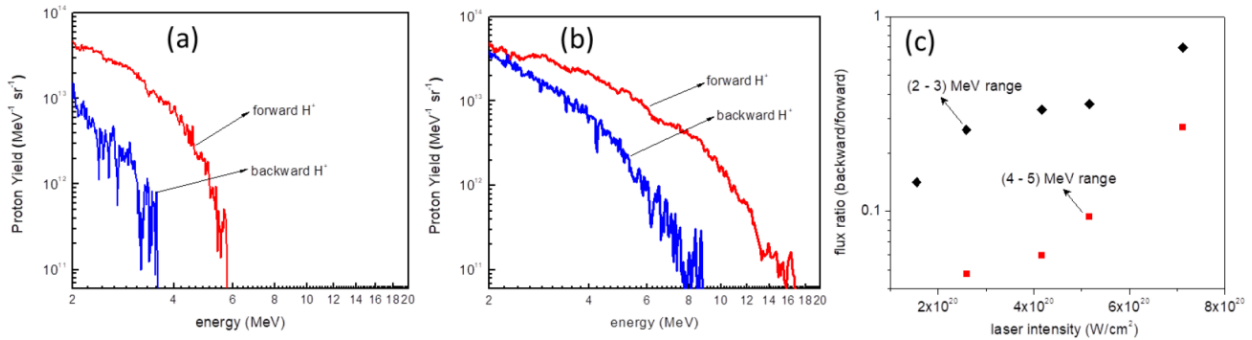

Fig. 3. Flux comparison of forward and backward protons. Kinetic energy spectrum of forward and backward accelerated protons for intensity of a) $1.5\times10^{20}$ W/cm$^2$ b) $7.2\times10^{20}$ W/cm$^2$. c) Variation of backward and forward proton flux ratio in the energy range of (2.0 - 3.0) MeV and (4.0 - 5.0) MeV

At higher laser intensity (Fig. 3b)), the difference between forward and backward proton flux is reduced. The ratio of backward and forward proton flux in two energy intervals of (2.5 - 3.5) MeV and (4.0 -5.0) MeV when laser intensity is increased by factor of 5 is shown in Fig. 3c). For both energy intervals, the flux ratio gradually increases with laser intensity. In the energy range of (2 - 3) MeV, the flux ratio rises close to unit at the highest laser intensity, whereas for the (4 - 5) MeV range, the ratio increases up to 0.3. This indicates that the ion flux along both side of the target gradually equalises with increase of laser intensity.

## 4. Intensity scaling of forward and backward accelerated ions

The ion energy scaling with laser intensity is a crucial parameter for determining the mechanism of acceleration process. To establish the scaling law, ion acceleration from an *Al* target of a wide range of thicknesses (0.2 - 6.0 µm) and laser intensities ($2\times10^{20}$ W/cm$^2$ - $1.2\times10^{21}$ W/cm$^2$) was investigated. The proton energy scaling dependent on laser intensity in the forward and backward directions is shown in the Fig. 4 a) and b), respectively. For the forward accelerated protons no significant target thickness dependence was observed, indicating that due to high level of laser contrast (~$10^{-10}$), even the thinnest targets (0.2 µm) is not affected by the pre-plasma [11]. The maximum proton energy observed in forward direction at laser-target parameters scan was nearly 30 MeV. The proton energy data, was fitted with a polynomial power law and shows proton energy scaling with laser intensity as $E_{forward} \propto I^{0.8}$ (thick gray line in Fig. 4a)). Fig. 4b) summarizes similar results obtained in the backward direction. All thicknesses of the targets, except 0.2 µm thick *Al* foil, show almost similar dependence on laser intensity with maximum proton energy nearly 18 MeV. A polynomial fitting to the data results in the intensity scaling of $E_{backward} \propto I^{0.6}$ (thick gray line in Fig. 4b)). The observation of different intensity scaling for protons, along two sides of the target surface, indicates that the acceleration mechanism is different in the forward and backward direction [5].

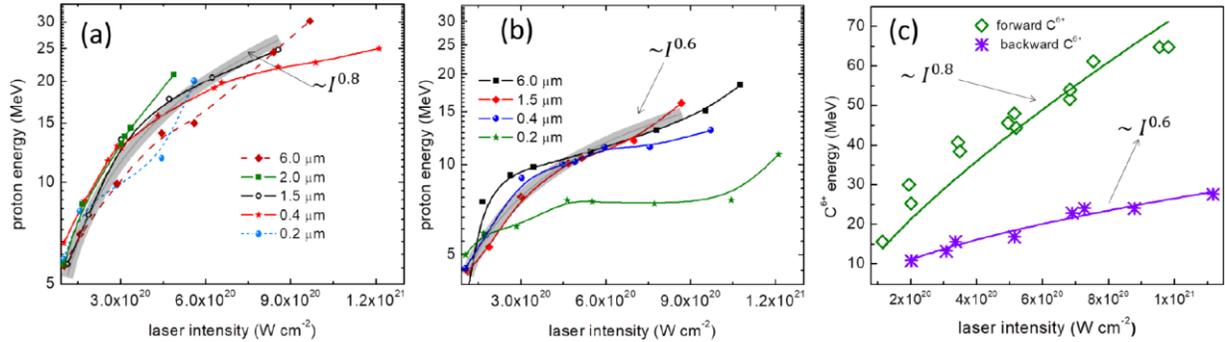

Fig. 4. Intensity scaling of proton and Carbon ions. Intensity dependent proton cut-off energy for different thickness of Al foil target is plotted for a) forward and b) backward direction. The thick grey line is the fitted curve on 1.5 µm data, representing intensity dependence of proton energies a) $E_{forward} \propto I^{0.8}$ and (b) $E_{backward} \propto I^{0.6}$ for forward and backward accelerated proton, respectively. c) Intensity dependent $C^{6+}$ cut-off energy is plotted for forward and backward direction. The solid lines are the fitted curve, representing intensity dependence of ion energies $E_{forward} \propto I^{0.8}$ and $E_{backward} \propto I^{0.6}$ for forward and backward $C^{6+}$, respectively.

The ion energy scaling with laser intensity in the forward and backward direction was also examined for carbon ion. Fig. 4c) shows the intensity scaling of $C^{6+}$ ion energies when 0.4 µm *Al* target was irradiated. The observed maximum $C^{6+}$ ion energy in forward and backward directions is nearly 65 MeV and 28 MeV, respectively. The fitted data show intensity scaling $E_{forward} \propto I^{0.8}$ and $E_{backward} \propto I^{0.6}$ for forward and backward $C^{6+}$ accelerated ions, respectively. Interestingly, the intensity scaling of $C^{6+}$ ions is similar to that observed for protons (Fig. 4 a) and b)).

## 5. Proton energy scaling: laser focal spot defocusing vs laser energy scan

The maximum energy of proton is associated to the laser intensity. However, laser intensity has three parameters, the laser energy, the laser focal size and the laser pulse duration, and the laser intensity change can be obtained by changing any of them. By fixing the laser pulse duration (30 fs), which can be linked with the time scales of acceleration processes, it is interesting to explore to which laser parameter change the proton energy scaling is more sensitive in changing the laser intensity [12]. Fig. 5 shows when 1.5 µm Al foil is irradiated to laser intensity varying in the range of $10^{19}$ W/cm$^2$ to $10^{21}$ W/cm$^2$. The maximum energy of forward and backward accelerated proton dependent on laser intensity, obtained by variation of laser focal spot in the target at constant laser energy, is shown in Fig. 5a).

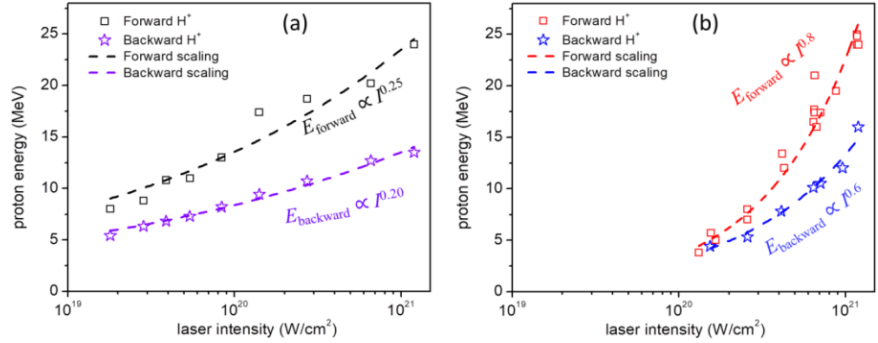

Fig.5. Proton intensity scaling with changing laser focal spot vs laser energy. Intensity dependent proton maximum energy scaling obtained under a) variation of the laser focal spot and b) by changing laser energy.

This was experimentally achieved by placing foil target at different positions (maximum displacement ~ 300 µm) from the laser focal plane. In the scanned intensity range of $10^{19}$ W/cm$^2$ to $10^{21}$ W/cm$^2$, the proton energy scaling is best fitted to $E_{forward} \propto I^{0.25}$ and $E_{backward} \propto I^{0.20}$, for the forward and backward protons, respectively. A similar comparison of proton energy is shown in Fig. 5b) for laser intensity variation as a function of laser energy at a fixed target position in the focal plane. Here, for the much shorter scanned intensity range of $10^{20}$ W/cm$^2$ to $10^{21}$ W/cm$^2$, the forward and backward proton energy scaling is observed to be $E_{forward} \propto I^{0.8}$ and $E_{backward} \propto I^{0.6}$, respectively. From these two comparative laser intensity scans, by either a change in laser energy or by laser focal spot conditions, the proton energy scaling is much faster with the changing the laser energy rather than the laser focal spot. This observation show that the goal to achieve proton beam with energy in the 100 MeV range can be pursued much faster by increasing the laser energy rather than achieving the tighter laser focusing conditions with small $f$-number optics [13].

## 6. Conclusions

In conclusion, experimental results of ion acceleration by petawatt femtosecond laser solid interaction have been presented. Irradiation of micrometer thick (0.2 µm - 6.0 µm) Al foils by 30 fs laser, in the intensity range of 8×$10^{19}$ W/cm$^2$ - $10^{21}$ W/cm$^2$, drives the acceleration of ions along the target normal rear and front side. The maximum of proton and carbon ion energies showing fast intensity scaling as $I^{0.8}$ along the forward direction and in backward direction as $I^{0.6}$. It is found that the ratio of maximum ion energy along the forward and backward direction to be constant for different target thicknesses and laser intensities. The proton energy scaling with laser intensity is examined by either changing the laser focal spot or by changing the laser energy. The observation of faster proton energy scaling with the changing of laser energy emphasis that the goal to achieve proton beam with energy in the 100 MeV range can be pursued much faster by increasing the laser energy rather than by achieving the tighter laser focusing conditions.

These investigations may help to find basic acceleration scheme for future applied efforts to validate advance modelling and improve predictions.

**Acknowledgement**

This work was performed under the ELI-ALPS Project (No. GINOP-2.3.6–15-2015–00001), which was supported by European Union and co-financed by the European Regional Development fund. The authors acknowledge funding from the Institute for Basic Science (IBS) under IBS-R012-D1, from EPSRC, through Grant Nos. EP/J002550/1, EP/L002221/1, EP/K022415/1, EP/J500094/1 and from Russian Foundation for Basic Research through Grant Nos. 15-02-03042 and 16-02-00088.